\documentclass[aps,twocolumn,prl,showpacs]{revtex4}
\usepackage{psfig}

\begin{document}

\title{Gaussian transformations and  distillation of entangled
Gaussian states}
\author{Jarom\'{\i}r Fiur\'{a}\v{s}ek}
\affiliation{Department of Optics, Palack\'{y} University, 17. listopadu 50,
77200 Olomouc, Czech Republic}

\begin{abstract}

We prove that it is impossible to  distill more entanglement from a
single copy of a two-mode bipartite entangled Gaussian state
via LOCC Gaussian operations. More generally, we show
that any hypothetical distillation protocol for Gaussian states
involving only Gaussian operations would be a deterministic protocol.
Finally, we argue that the protocol considered by Eisert {\em et al.}
[quant-ph/0204052] is the optimum Gaussian distillation protocol for
two copies of entangled Gaussian states.

\end{abstract}

\maketitle

Quantum entanglement is a key ingredient of many protocols for
quantum information processing such as quantum teleportation
\cite{Bennett93} or quantum cryptography \cite{Ekert91}.
Usually, the entangled particles are distributed among two  distant parties
traditionally called Alice and Bob. In practice, the transmission
channel used for this distribution is always noisy and imperfect,
which prevents Alice and Bob from sharing a maximally entangled state even
if Alice can prepare such state locally in her lab.
Fortunately, the errors introduced by noisy quantum channels can be overcome
by the so-called entanglement distillation protocols, by which Alice and Bob
can extract from a large number of weakly entangled  mixed states
a smaller number of highly entangled almost pure states
\cite{Bennett96,Deutsch96}.

Recently, a great deal of attention has been devoted to the quantum
information processing with continuous quantum variables and
continuous-variable analogues of various protocols developed
originally in the framework of discrete quantum variables have been
established. Remarkably, linear optics, parametric
amplifiers, and homodyne detectors suffice for implementation of
many of these protocols including continuous-variable teleportation
\cite{Braunstein98}, cryptography \cite{Ralph00},
and cloning \cite{Braunstein01}. However, one
important protocol missing in our toolbox is a feasible distillation
protocol for continuous variables. We are particularly interested in
distillation protocols for entangled Gaussian states because these
states can easily be generated in the laboratory. The protocols suggested
so far involve rather complicated nonlinear transformations such as subtraction
of a single photon \cite{Opatrny00} or quantum non-demolition measurement
of total photon number in several modes \cite{Duan00}.
It would be of great help to have a distillation protocol
for continuous variables that could be implemented with linear optics
and which would distill Gaussian entangled states. However, no such
protocol is currently known and it is an open question whether such
distillation protocol exists at all.

In the present paper we attempt to shed some light on this issue
making use of the formalism of Gaussian completely positive (CP) maps
\cite{Lindblad00,Eisert01}.
These maps represent all transformations that can be carried out with
the help of passive and active linear optical elements, homodyne
detectors, and auxiliary optical modes prepared initially in
Gaussian states. These transformations may be deterministic or
probabilistic. In the latter case we accept or reject the output state
in dependence on the output of a quantum measurement (with some Gaussian
probability distribution). We shall consider an arbitrary bipartite
probabilistic Gaussian operation which can be implemented with the help
of local operations and classical communication (LOCC).
We shall prove that for input bipartite Gaussian states
it is always possible to construct a deterministic LOCC
Gaussian transformation that yields the same output state (for a fixed input)
as a given probabilistic LOCC Gaussian transformation.
This implies that it is impossible to distill more entanglement from
a single copy of entangled Gaussian state by means of Gaussian
operations. This should be contrasted with distillation protocols for
a single copy of two-qubit entangled state where the LOCC operations may
in some cases allow to extract more entanglement \cite{Kent99}.
In particular, any pure entangled two-qubit state can be transformed with
certain probability via LOCC operations onto maximally entangled Bell state.
Furthermore, our results imply that any hypothetical Gaussian
distillation protocol optimized for given shared entangled Gaussian states
would be a deterministic protocol and we shall find a generic structure
of this optimum protocol.
A version of this optimum protocol where Alice and Bob share two
identical copies of Gaussian state with symmetric covariance matrix
has been considered by Eisert {\em et al.} \cite{Eisert02}
who employed the log-negativity as the entanglement measure and proved
that it is impossible to distill entanglement via this protocol.
These findings thus strongly support the conjecture that it is impossible
to distill entangled Gaussian states via Gaussian operations and that some
nonlinearity is necessary.

We shall extensively exploit the Jamiolkowski isomorphism
\cite{Jamiolkowski72}
between completely positive maps $\cal{M}$ and positive-semidefinite operators
(bipartite quantum states) $\chi$ on tensor product of input and output
Hilbert spaces ${\cal{H}}{\otimes \cal{K}}$. In terms
of the operator $\chi \geq 0$ the relation between input and output density
matrices can be written as a partial trace over the input space,
\begin{equation}
\rho_{\rm out}= {\rm Tr}_{\rm in}[\chi \, \rho_{\rm in}^T
\otimes\openone_{\rm out}],
\label{INOUT}
\end{equation}
where $T$ stands for the transposition in some fixed basis and
$\openone_{\rm out}$ denotes an identity operator on output space.
The operator  $\chi$
can be obtained from a maximally entangled state on
${\cal{H}}^{\otimes 2}$ ,
$|\psi\rangle= \sum_{j=1}^d |j\rangle_1|j\rangle_2$, ($d=\rm dim\cal{H}$)
if the map $\cal{M}$ is applied to one part of this state,
\begin{equation}
\chi={\cal{M}} \otimes {\cal{I}}\left[|\psi\rangle\langle\psi|\right].
\label{DEFCHI}
\end{equation}
Here $\cal{I}$ stands for the identity transformation.

In continuous variable systems, we deal with infinite dimensional
Hilbert spaces and the maximally entangled state $|\psi\rangle$ becomes
a tensor product of $N_{\rm in}$ (unphysical)
two-mode infinitely squeezed vacuum states,
where $N_{\rm in}$ is the number of input modes. Gaussian completely
positive maps are defined as maps which transform Gaussian states into
Gaussian states. Gaussian CP maps are thus isomorphic to bipartite Gaussian
quantum states $\chi$. Now any Gaussian state $\chi$ is completely characterized
by the first and second moments: mean values of quadratures and a covariance
matrix $\Gamma$.
Define vector of quadratures
$\vec{r}=(x_1,p_1,\ldots,x_N,p_N)^T$ where $N$ is the
total number of input$+$output modes. The elements of matrix $\Gamma$ are
defined as
$\Gamma_{ij}= \langle \Delta r_i \Delta r_j \rangle + \langle \Delta r_j
\Delta r_i \rangle,$
where $\Delta r_i=r_i- \langle r_i\rangle$.
Nonzero mean values of the quadratures of the Gaussian state $\chi$
representing a Gaussian CP map indicate that this map involves
certain displacements. However, these operations can be performed locally and
are therefore irrelevant for the entanglement properties and
can be omitted. Thus we can assume that  $\langle r_i\rangle=0 $ and the
CP map $\chi$ is fully described by the covariance matrix $\Gamma$.
It is convenient to split the matrix $\Gamma$ into input and output parts and
write
\begin{equation}
\Gamma=\left(
\begin{array}{cc}
A & C \\
C^T & B
\end{array}
\right),
\label{GAMMADECOMPOSED}
\end{equation}
where $A$ stands for the covariance matrix of the ``input'' modes, $B$ is
the covariance matrix of ``output'' modes and $C$ contains the input-output
correlations.  The input-output transformation (\ref{INOUT}) rewritten in
terms of the Wigner functions reads
\begin{equation}
W_{\rm out}(\vec{r}_{\rm out})= (2\pi)^{N_{\rm in}}\int_{-\infty}^\infty
W_{\chi}(\vec{r}_{\rm in},
\vec{r}_{\rm out}) W_{\rm in}( R \vec{r}_{\rm in} ) \, d \vec{r}_{\rm in},
\label{WINOUT}
\end{equation}
where  $R={\rm diag}(1,-1,1,-1,\ldots,1,-1)$
is a diagonal matrix which represents the transposition in
phase space ($x_j\rightarrow x_j$, $p_j\rightarrow-p_j$).
It is convenient to deal with characteristic functions which are Fourier
transforms of the Wigner functions,
\begin{equation}
{\cal{C}}(\vec{q})= \int_{-\infty}^\infty
W(\vec{r}) \exp\left( i \vec{r} \cdot \vec{q} \right) \, d \vec{r}.
\label{CHARFUNCTION}
\end{equation}
On expressing all Wigner functions in terms of the characteristic
functions, we obtain from Eq. (\ref{WINOUT}),
\begin{equation}
{\cal{C}}(\vec{q}_{\rm out})= (2\pi)^{-N_{\rm in}} \int_{-\infty}^\infty
{\cal{C}}_{\chi}(\vec{q}_{\rm in},
\vec{q}_{\rm out}) {\cal{C}}_{\rm in}(-R\vec{q}_{\rm in}) \,d \vec{q}_{\rm in}.
\label{CHARINOUT}
\end{equation}
Assuming input Gaussian state with covariance matrix $\Gamma_{\rm in}$,
${\cal{C}}_{\rm in}(\vec{q})= \exp\left(- \frac{1}{4}
\vec{q}^T \Gamma_{\rm in} \vec{q} \right),
$
we find that the the output state is also Gaussian with covariance
matrix given by
\begin{equation}
\Gamma_{\rm out}= B- C^T(A+R\Gamma_{\rm in}R)^{-1}C.
\label{GAMMAINOUT}
\end{equation}

We now prove a very important feature of Gaussian CP maps. It holds that
for every input Gaussian state and  a probabilistic (trace-decreasing)
LOCC Gaussian CP map there exists a deterministic (trace-preserving)
LOCC Gaussian CP map which transforms the input state into the output state
with the covariance matrix (\ref{GAMMAINOUT}).
The explicit construction of the trace-preserving map is inspired by
recent works on the possibility of storing quantum dynamics in quantum states
\cite{Dur01,Hillery02}.
The basic strategy is to encode the transformation into a bipartite
state $\chi$ which then serves as a quantum channel in the teleportation.
In this way, the desired transformation is carried out with certain
probability depending on the dimension of the Hilbert space.

The continuous-variable analogue of this scheme goes as follows.
We prepare a Gaussian state $\chi$  with covariance matrix $\Gamma$
given by Eq. (\ref{GAMMADECOMPOSED}) and
carry out a Bell measurement  on the input state and the input modes of the
state $\chi$. This measurement is performed separately for each
corresponding pair of modes and consists of measuring the difference of
$x$ quadratures and sum of $p$ quadratures by means of homodyne
detectors.
Let the vectors $\vec{x}_{d}$ and $\vec{p}_{d}$ contain the measurement
outcomes for $x$ and $p$ quadratures, respectively.
The (non-normalized) Wigner function of the output modes conditioned
on the measurement outcome $\vec{r}_d$ reads
\begin{eqnarray*}
W_{\rm out}(\vec{r}_{\rm out}| \vec{r}_{d})&=&
\int_{-\infty}^\infty W_{\chi}(\vec{r}_{\rm in},\vec{r}_{\rm out})
W_{\rm in}(\vec{r}) \delta(\vec{x}_{\rm in}-\vec{x}-\vec{x}_{d})
\nonumber \\ &&
\quad\times\delta(\vec{p}_{\rm in}+\vec{p}-\vec{p}_{d}) \,
d \vec{x}_{\rm in} \, d \vec{p}_{\rm in} \, d \vec{x} \, d \vec{p}.
\nonumber
\label{WOUTCOND}
\end{eqnarray*}
Consider input Gaussian state. It turns out that the covariance matrix of the
output state is given by Eq. (\ref{GAMMAINOUT}) and does not depend
on the measured quadratures $\vec{x}_d$ and $\vec{p}_d$.
However, the output state is displaced by
\begin{equation}
\vec{r}_{\rm cond}= C^T (A+R\Gamma_{\rm in}R)^{-1}  \vec{r}_{d}.
\label{RDIS}
\end{equation}
If we know the input state and the transformation, then we can calculate
$\vec{r}_{\rm cond}$ for given detected quadratures
$\vec{x}_{d}$, $\vec{p}_d$ and
by means of suitable displacement transformation applied to output
state we can always set the coherent signal in  the output state to
zero.  In this way we obtain in a deterministic manner an output  state
which has the covariance matrix (\ref{GAMMAINOUT}).
This procedure is essentially the Braunstein-Kimble scheme for teleportation
of continuous variables \cite{Braunstein98}. We should note here that many
Gaussian CP maps, in particular all trace-preserving maps,
are represented by unphysical
(infinitely squeezed) states $\chi$. However, we can approximate
such unphysical state by a physical finitely squeezed state with
an arbitrarily high accuracy and thus also approximate the transformation
(\ref{GAMMAINOUT}) with arbitrarily  high precision.

\begin{figure}
\centerline{\psfig{figure=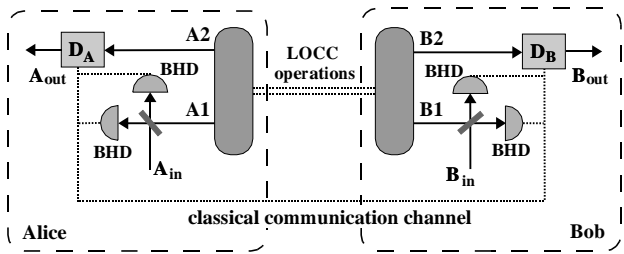,width=0.95\linewidth}}
\caption{Setup for implementation of deterministic LOCC Gaussian CP map
that is for certain input Gaussian state equivalent to a given
probabilistic LOCC Gaussian CP map.}
\end{figure}

Let us now turn our attention to the LOCC Gaussian CP maps.
Obviously, every LOCC Gaussian map ${\cal{M}}_{\rm LOCC}$ is isomorphic to
a Gaussian state $\chi$ which is separable with respect to Alice and Bob.
(Note that according to Eq. (\ref{DEFCHI}) Alice and Bob can prepare
the state $\chi$ in their labs via LOCC operations.)
A scheme for deterministic implementation of any LOCC Gaussian CP map
(for known input Gaussian state) is shown in Fig. 1 for the simplest case when
there is a single input and a single output mode on each side.
By means of LOCC operations, Alice and Bob prepare the four-mode state
$\chi$ representing the CP map ($A1$ and $B1$ are input modes and
 $A2$ and $B2$ are output modes).
Alice mixes her part of the input state in mode $A_{\rm in}$
with $A1$ on balanced beam  splitter and measures $x_{A1}-x_{A\rm in}$
and $p_{A\rm in}+p_{A1}$ by means of two balanced homodyne
detectors. Bob performs the same operations with his modes $B1$ and
$B_{\rm in}$. Alice and Bob exchange the results of their measurements
via classical communication channel and apply appropriate local displacement
transformations to the modes $A2$ and $B2$ thereby producing the two-mode
output state $A_{\rm out}$ and $B_{\rm out}$.
Notice that the protocol
works only for Gaussian states. If the input state is not Gaussian, than
it may happen that some trace-decreasing LOCC Gaussian CP maps will yield
outputs that cannot be obtained with any trace-preserving LOCC
Gaussian CP map.

A very important implication concerning distillation protocols is that
we cannot distill more entanglement from a single copy of two-mode bipartite
entangled Gaussian state by means of LOCC Gaussian operations.
This follows from the fact that any probabilistic LOCC Gaussian operation
can be replaced by a deterministic one which yields the same output
(for a given input Gaussian state). However, any reasonable measure of
entanglement must be non-increasing under deterministic LOCC operations.
Here we note that this impossibility was noted in a paper by
Parker {\em et al.} \cite{Parker00}, who considered distillation via
continuous-variable entanglement swapping. While this procedure works for
Schr\"{o}dinger cat states, it fails for two-mode squeezed vacuum.
Parker {\em et al.} also noticed that in the latter case,
the probabilistic Gaussian transformation becomes deterministic.
Our arguments show that this is a general feature of Gaussian states and
Gaussian CP maps.

Let us have a more detailed look at the structure of bipartite LOCC
Gaussian CP maps. It was shown by  Werner and Wolf \cite{Werner01}
that a bipartite Gaussian state is separable if and only if it can be
transformed via local symplectic transformations into state with positive
Glauber-Sudarshan representation, i.e. a state which is a convex mixture
of coherent states and is not squeezed. We can thus write
\begin{equation}
\chi_{\rm LOCC}= \int_{-\infty}^\infty P(\alpha,\beta) S_A|
\alpha\rangle\langle \alpha|S_A^\dagger \otimes
 S_{B}|\beta\rangle\langle \beta| S_{B}^\dagger  \, d\alpha \, d\beta,
 \label{CHILOCC}
\end{equation}
where $P(\alpha,\beta)\geq 0$ is classical Gaussian probability distribution
and $|\alpha\rangle$, $|\beta\rangle$ denote  (multimode) coherent states
of Alice's and Bob's modes.
We have seen that it is impossible to distill a single copy of a Gaussian
state by means of LOCC Gaussian transformations.
What if Alice and Bob posses several copies?
Assume that Alice and Bob apply the LOCC Gaussian map (\ref{CHILOCC})
to their states. This distillation map  takes all copies as an input and
yields a single copy of two-mode state shared by Alice and Bob as the
output. The output state is a mixture of states
with identical covariance matrices and varying displacements.
In terms of Wigner functions, we can write
\begin{equation}
W_{\rm out}(\vec{r})= \int_{-\infty}^\infty
\tilde{P}(\alpha,\beta)W(\vec{r}-\vec{r}_d(\alpha,\beta)) \,
d\alpha \, d\beta,
\label{WMIXTURE}
\end{equation}
where  $\vec{r}_{d}(\alpha,\beta)$ is the displacement and
$\tilde{P}(\alpha,\beta)\geq 0$. However, all states with Wigner functions
$W(\vec{r}-\vec{r}_d)$ exhibit the same entanglement, because entanglement
depends only on the covariance matrix and not on the displacement.
Hence it is always optimum to choose an LOCC CP map which is represented
by a {\em pure} Gaussian state.
An analogous situation arises in distillation of a
single pair of entangled qubits where it is optimum  to apply local filtering
operation \cite{Kent99} (this is a trace-decreasing CP map whose
Kraus decomposition contains only one term and the CP map
is thus represented by a pure state).

Since we assume that we know the state that we want to distill,
we can transform any LOCC trace-decreasing map onto trace-preserving
map hence the optimum protocol will be deterministic.
Consider now the simplest case when Alice and Bob share two pairs
of entangled Gaussian states. The transformation on Alice's side
is represented by a pure three-mode Gaussian state, which can be
obtained from three-mode vacuum via some three-mode symplectic
transformation. This three-mode state splits into two input modes
and one output mode. By means of ``local'' symplectic transformations
on input and output modes, we can transform the covariance matrix of
this state into the form
\begin{equation}
\Gamma= \left(
\begin{array}{cccccc}
a & 0 & 0 & 0 & d_1 & 0   \\
0 & a & 0 & 0 & 0   & d_2 \\
0 & 0 & b & 0 & e_1 & 0   \\
0 & 0 & 0 & b & e_3   & e_2 \\
d_1 & 0 & e_1 & e_3 & c & 0 \\
0 & d_2 & 0 & e_2 & 0 & c
\end{array}
\right).
\label{GAMMAABC}
\end{equation}
The reduced density matrix of each input and output mode is density
matrix of thermal state and there are no correlations between the
two input modes. Since the whole three-mode state is pure, the
density matrix of the two input states  has the same eigenvalues
as the reduced density matrix of the output state. This is possible only
if  one of the input modes is in pure vacuum state. This leads to
$a=c$, $b=1$, $e_1=e_2=e_3=0$. One of the input modes is effectively
decoupled and we end up with pure  two-mode squeezed vacuum
state. We thus have a very appealing and intuitive picture:
any pure three-mode state can be prepared if we start from (suitably
chosen) two-mode squeezed vacuum state and a vacuum state and apply
single-mode symplectic transformation to the output mode and two-mode
symplectic transformation $S_i$ to two input modes.
We may thus write the transformation on Alice's side in the form
\begin{equation}
\chi_A= S_{i}  (\chi_{A0}\otimes |0\rangle \langle 0|) S_i^\dagger,
\label{CHIA}
\end{equation}
where $\chi_{A0}$ is an operator on Hilbert space of two modes
(one input and one output). To see what are the implications,
we insert this expression into formula (\ref{INOUT}) and for the moment do not
consider Bob's states. We get
\begin{eqnarray}
\rho_{\rm out}
&=&{\rm Tr}_{\rm in} [\chi_{A0}\otimes |0\rangle\langle 0| S_i^\dagger
\rho_{\rm in}^T S_{i} \otimes \openone_{\rm out}]
\nonumber \\
&=&{\rm Tr}_{\rm in} [\chi_{A0}\otimes |0\rangle\langle 0|
(S_i^T \rho_{\rm in} S_{i}^{\ast})^T \otimes \openone_{\rm out}].
\label{RHOOUTCHIA}
\end{eqnarray}
From this formula we can see that the transformation reduces to the
following three steps: (i) apply symplectic transformation $S_i^T$
to the input two-mode state. (ii) project the second mode onto vacuum
state. (iii) Apply a CP map $\chi_{A0}$ to the first mode.
The transformation on Bob's side has the same structure.
\begin{figure}[t]
\centerline{\psfig{figure=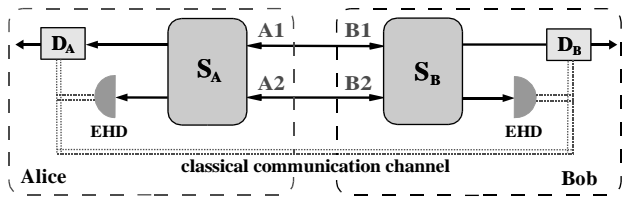,width=0.95\linewidth}}
\caption{Optimal distillation protocol for two copies of two-mode
entangled Gaussian states.}
\end{figure}
This protocol can be further simplified. After Alice and Bob
project one of modes onto vacuum state, they possess only a single mode
each. The application of the local maps $\chi_{A0}$ ($\chi_{B0}$)
cannot increase entanglement, because we have shown that it is
impossible to distill a single copy of two-mode entangled Gaussian state
by means of LOCC Gaussian operations. Thus  we need not consider the
transformations $\chi_{A0}$ ($\chi_{B0}$) and the resulting optimum
simplified distillation scheme is shown in Fig. 2.
Alice and Bob share two pairs of Gaussian states. Alice's and Bob's
modes are labeled as $A1$, $A2$  and $B1$, $B2$, respectively.
Both Alice and Bob locally apply some two-mode symplectic
transformations $S_A$ and $S_B$ to their modes. Subsequently they both
feed the modes  $A2$ and $B2$  to eight-port homodyne detectors (EHD)
thereby projecting them into coherent states $|\alpha\rangle$
and $|\beta\rangle$.
Finally, they exchange the results of their measurements
and displace appropriately the output states. This scheme is deterministic and
represents the optimum Gaussian distillation protocol for Gaussian
states. Eisert {\em et al.} proved that it is impossible to distill
entanglement from two identical copies of two-mode Gaussian state with
symmetrical covariance matrix via this protocol \cite{Eisert02}.
All these results strongly support the conjecture that it is impossible
to distill entangled Gaussian states with Gaussian operations.

{\em Note added:} After this work was completed, I learned that
Giedke and Cirac \cite{Giedke02} have also investigated the properties of
trace-decreasing Gaussian CP maps and they independently obtained similar
results. Moreover, they proved that the distillation of Gaussian states
with Gaussian operations is impossible for an arbitrary number of
modes per site.

I would like to thank J. Eisert, P. van Loock, G. Giedke, N. Cerf,
R. Filip, and L. Mi\v{s}ta, Jr. for discussions.
This work was supported by Grant No LN00A015
of the Czech Ministry of Education and by the EU grant under
QIPC, project IST-1999-13071 (QUICOV).

\end{document}